\documentclass[twoside]{article}
\usepackage{fleqn,espcrc2}

\newcommand{\AmS}{{\protect\the\textfont2
  A\kern-.1667em\lower.5ex\hbox{M}\kern-.125emS}}

\title{The lagrangian description of representations
       of the Poincare group}

\author{{\v{C}. Burd\'{\i}k}\address{Department of Mathematics, Czech
Technical
University, \\Trojanova 13, 120 00 Prague 2}\address[MCSD]
{Bogoliubov Laboratory of Theoretical Physics,
JINR,\\
Dubna 141980, Russia}%
        \thanks{burdik@thsun1.jinr.ru}%
        \thanks{Talk given  at the D.V. Volkov
Memorial Conference ``Supersymmetry and Quantum Field Theory'', July 25-30, 2000,
Kharkov},
A. Pashnev\addressmark[MCSD]
\thanks{pashnev@thsun1.jinr.ru},
    and M. Tsulaia\addressmark[MCSD]\address{The Abdus Salam International
Centre for Theoretical Physics,\\
 Trieste 34014, Italy}\address{
Andronikashvili Institute of Physics,\\
 Tbilisi 380077, Georgia}%
\thanks{tsulaia@thsun1.jinr.ru}}

\begin{document}

\begin{abstract}
The construction of lagrangians describing the various representations
of the Poincare group is given in terms of the BRST approach.
\vspace{1pc}
\end{abstract}

\maketitle

\section{THE DESCRIPTION OF CONSTRAINTS}

It is well known that the particles with the value of spin
more than two arise naturally when quantizing such classical objects as
the relativistic oscillator \cite{BD}, string \cite{GSW} or discrete
string
\cite{GP,FI}. The challenging
problem  for these kinds of theories is to construct the lagrangian
description both for free and for interacting particles with
the higher spins.

In order to  kill ghosts and all other superfluous representations
the field theoretical lagrangians, describing irreducible Poincare
representations must possess some gauge invariance.
Along with the basic fields such lagrangians in general include
additional ones.
The role of these fields is to single out the irreducible
representation of the Poincare group.
Some of them are auxiliary, others can be gauged away.
After a gauge fixing and solving the equations of motion for auxiliary
fields one is left  with the only essential field, describing the
irreducible representation of the
Poincare group. In general this field
 corresponds to the Young table with $k$ rows
and is described by
$\Phi^{(k)}_{\mu_1\mu_2\cdots\mu_{n_1}, \nu_1\nu_2\cdots\nu_{n_2},
\cdots,\rho_1\rho_2\cdots\rho_{n_k}}(x)$ which is the
  $n_1+n_2+\cdots+n_k$ rank tensor field
symmetrical with respect to the
permutations of each type of indices.
In addition, this field is subject to
 the mass shell condition and
transversality conditions
for each type of indices.
 Further, all traces of the basic field must vanish.
The correspondence with a given Young table implies, that
after symmetrization
of all indices from  $i$ - th row with  one index from $j$ - th row
$(i<j)$
the basic field vanishes,
for example
\begin{equation}
\label{sym}
\Phi^{(k)}_{\bf \{\mu_1\mu_2\cdots\mu_{n_1}, \nu_1\}\nu_2\cdots\nu_{n_2},
\cdots,\rho_1\rho_2\cdots\rho_{n_k}}(x)=0.
\end{equation}

To describe all irreducible representations of the Poincare group
simultaneously it is convenient to
introduce an auxiliary Fock space generated by
the creation and annihilation
operators $a^{i+}_\mu,a^j_\mu$
with Lorentz index $\mu =0,1,2,...,D-1$ and additional internal
index $i=1,2,...,k$. These operators satisfy the following
commutation relations
\begin{equation} \label{osc}
\left[ a^i_\mu,a^{j+}_\nu \right] =-g_{\mu \nu}\delta^{ij},
\end{equation}
\begin{equation}
g_{\mu \nu}=diag(1,-1,-1,...,-1),
\end{equation}
where $\delta^{ij}$ is usual Cronecker symbol.

The general state of the Fock space
depends on the space - time coordinates $x_\mu$
\begin{eqnarray}
&&|\Phi\rangle =\sum
\Phi^{(k)}_{\mu_1\mu_2\cdots\mu_{n_1},
\cdots,\rho_1\rho_2\cdots\rho_{n_k}}(x) \times \nonumber \\
&&a^{1+}_{\mu_1}a^{1+}_{\mu_2}\cdots a^{1+}_{\mu_{n_1}}
\cdots
a^{k+}_{\rho_1}a^{k+}_{\rho_2}\cdots a^{k+}_{\rho_{n_k}}
|0\rangle
\end{eqnarray}
and the components
$\Phi^{(k)}_{\mu_1\mu_2\cdots\mu_{n_1},
\cdots,\rho_1\rho_2\cdots\rho_{n_k}}(x)$
are automatically symmetrical under the permutations of indices of
the same type.
The norm of states in this Fock space is not positively definite due to
the minus sign in the commutation relation (\ref{osc}) for the time
components of
the creation and annihilation operators.
The transversality conditions
for the components leading to the positively definite norm of physical
states
 are equivalent to the following constraints on the
physical vectors of the Fock space
\begin{equation}
\label{trans3}
L^i|\Phi\rangle=0,\quad  L^i= a^i_\mu p_\mu.
\end{equation}
These operators $ L^i$ along with their conjugates
$L^{i+}= a^{i+}_\mu p_\mu$
and mass shell operator $p^2_\mu$
form the following algebra with only nonvanishing commutator
\begin{equation}     \label{algebra1}
\left[L^i,L^{j+}\right]=-p^2_\mu\delta^{ij}.
\end{equation}
This simple algebra was considered in \cite{OS} in the framework of
the BRST approach, which automatically leads
to the appearance of all auxiliary fields in the lagrangian.
 Since the constraints are of the first class
the corresponding nilpotent BRST charge can be constructed
straightforwardly.
As a result
the description of mixed symmetry
(not only totally symmetric)
fields was obtained. However, all
these fields describe the reducible representations  of the Poincare group
 due to the absence of tracelessness and mixed symmetry conditions
in the initial system of
the constraints.

On the other hand, the same algebra
of operators arises
in the case of massive particles. The only difference is that the
right hand side of the relation (\ref{algebra1}) is now nonvanishing
operator. Moreover, this operator can have different eigenvalues
$p^2_\mu=m_n^2$ for
the different physical states.
Therefore the procedure of construction of the nilpotent BRST charge for
this simple system
of constraints is complicated , since now   constraints $L^{i \pm }$ are
of the second class.

The tracelessness conditions
 correspond in the  Fock space to the
constraints
\begin{equation}
\label{trace4}
L^{ij}|\Phi\rangle=0,
\end{equation}
with
$ L^{ij}=a^i_\mu a^j_\mu,$
for $i \neq j$ and
$ L^{ij}=\frac{1}{2} a^i_\mu a^j_\mu,$
for $i = j$,
while the mixed symmetry properties  follow from the
constraints
\begin{equation}
\label{sym2}
T^{ij}|\Phi\rangle=0, \quad \;\;i<j,
\end{equation}
with
$T^{ij}=a^{i+}_\mu a^j_\mu.$
The operators $L^{ij}, \;L^{ij+}$ ($i,j$ are arbitrary) and
$T^{ij}$,
$T^{ij+},\; (i\neq j)$, along with the additional operators
\begin{equation}
\label{Cartan}
H^i= -T^{ii}+\frac{D}{2}=-a^{i+}_\mu a^i_\mu+\frac{D}{2},
\end{equation}
 form the Lie algebra $SO(k+1,k)$.
The rank of this algebra is $k$ and corresponding Cartan subalgebra
contains all operators $H^i$. One can choose
the operators $L^{11}$ and $T^{i,i+1}$ as $k$ simple roots.
 The positive and negative
roots are, correspondingly, $L^{ij},\;T^{rs},\;(1\leq r<s\leq k)$
and $L^{ij+},\;T^{rs},\;(1\leq s<r\leq k)$. It means, that the
tracelessness and the mixed symmetry conditions  are equivalent
to annihilation of physical states in the total Fock space
by the positive roots of the Lie algebra $SO(k+1,k)$.
As it can be easily seen, the Cartan generators
(\ref{Cartan})  are strictly positive in the Fock space and
have to be excluded from the total set of constraints. Therefore
the standard BRST charge has to be modified for the given
realization of the  $SO(k+1,k)$ algebra.

In the present talk we give the lagrangian description of the massive
reducible representations of the Poincare group and  its irreducible
massless representations  with the corresponding Young
tableaux having two or one  rows. The later lagrangian can be
straightforwardly
reduced  to the lagrangian of  \cite{F}, describing the irreducible
massless higher spin fields
(see also \cite{PT2}).
Though in both cases the corresponding system of constraints
contains both the ones of the first and the second class, the nontrivial
structure of trilinear  ghost terms in the BRST charge makes it nilpotent.
The method used for these constructions can be generalized  for  arbitrary
representations of Poincare group as well.

\section{REDUCIBLE MASSIVE CASE}

To demonstrate the general method of BRST construction with
the second class constraints of the considered type we begin with the
simplest example of nonvanishing mass.
In this section we consider the system with constraints
\begin{equation} \nonumber
L^0 =-{p_{\mu}}^2+m^2,
\end{equation}
\begin{equation} \nonumber
L^1 = p_{\mu}a_{\mu},\;\;\; L^{1+}=p_{\mu}a_{\mu}^+,
\end{equation}
This system is
intermediate  between the systems in \cite{OS} and \cite{P} because
it describes the massive particle together with all its daughter
(due to the absence of the constraints $L^{\pm 11}$ in the theory are
present
particles with lower spins).

The commutation relation (\ref{algebra1})
(we are considering the case of one oscillator i.e., $i,j=1$)  means that
$L^{\pm 1}$ are the second class   constraints.
The simplest way to construct BRST - charge with corresponding constraints
 \cite{TH} -- \cite{PT1} is to consider massless case in $D+1$ dimensions
with constraints
\begin{equation} \nonumber
L^0=-p_{\mu}^2+p_D^2, \;\;\;\;\mu=0,1,...,D-1,
\end{equation}
\begin{equation} \nonumber
L^1=p_\mu a_\mu-p_Da_D,
\end{equation}
\begin{equation} \nonumber
L^{1+}=p_\mu a_\mu^+ -p_Da_D^+.
\end{equation}
Following to the standard procedure we
introduce additional set of anticommuting variables
$\eta_0,\eta_1,\eta_1^+$ having ghost number one and corresponding momenta
${\cal{P}}_0,{\cal{P}}_1^+,{\cal{P}}_1$ with commutation relations:
\begin{equation}
\{\eta_{0},{\cal{P}}_{0}\}=\{\eta_{1}, {\cal{P}}_{1}^+\}=
\{\eta_{1}^+,{\cal{P}}_{1}\}=1.
\end{equation}
and consider the total Fock space generated by creation operators
$a_\mu^+,a_D^+,\eta_{1}^+, \eta_{0}, {\cal{P}}_{1}^+$.
In terms of the nilpotent BRST - charge which corresponds to the given
system of constraints
\begin{equation}
Q =\eta_0  L^0 + \eta^+_{1} L^1 + \eta_{1}  L^{1+}
-\eta^{+}_{1}\eta_{1} {\cal{P}}_{0}
\end{equation}
the BRST invariant lagrangian can be written as follows
\begin{equation} \label{L11}
L=-\int d \eta_0 \langle\chi|Q|\chi\rangle.
\end{equation}

In order to describe the massive particle we fix the following
$x_D$
dependence of the Fock space vector $| {\chi} \rangle $
in (\ref{L11}):
\begin{equation}               \label{sub}
|{\chi}\rangle = U|\chi^{\prime}\rangle =
 e^{ ix_D m}|\chi^{\prime}\rangle.
\end{equation}
The result of the substitution of (\ref{sub}) in the expression
(\ref{L11})
is
\begin{equation} \label{L12}
L=-\int d \eta_0 \langle {\chi}^{\prime} |\tilde{Q}| {\chi}^{\prime}
\rangle,
\end{equation}
The new BRST - charge $\tilde{Q}$  is nilpotent due to
unitarity of the transformation $\tilde{Q}=U^{-1} Q U$. Our choice of
$x_D$- dependence in the exponent in (\ref{sub}) leads to the presence
in the BRST charge of the correct massive operator
$\tilde{L}_0=-p_\mu^2+m^2$
The same trick with the dimensional reduction will be applied in the
consequent sections to more complicated cases. It is exactly the heart
of the approach, which makes it possible to construct the
nilpotent BRST charge in the presence of the second class constraints.
The nilpotency of the BRST - charge evidently does not depend from the
unitary transformation of the type (\ref{sub}).

In order to be physical, the lagrangian  must have zero ghost number. It
means
that  the most general expressions for the vector $| \chi^{\prime}
\rangle$ is
\begin{equation} \label{vector}
|\chi^{\prime}\rangle = |S_1 \rangle+\eta_1^+{\cal{P}}_1^+|S_2\rangle,+
\eta_0 {\cal{P}}_1^+ |S_3\rangle,
\end{equation}
with vectors $|S_i\rangle$ having ghost number zero and depending only
on bosonic creation operators $a_\mu^+,a_D^+$
\begin{equation}\label{RA}
|S_i\rangle\!=\!\sum \phi^{n}_{\mu_1,\mu_2,...\mu_n}(x)
a_{\mu_1}^+ a_{\mu_2}^+ ...a_{\mu_n}^+(a_D^+)^{n}|0\rangle.
\end{equation}

The nilpotency of the BRST - charge leads to the invariance of the
lagrangian (\ref{L12}) under the following transformations
\begin{equation}\label{TR}
\delta |\chi^{\prime} \rangle = \tilde Q |\Lambda \rangle.
\end{equation}
The parameter of transformation must have ghost number $-1$ and
can be written as  $|\Lambda \rangle = {\cal{P}}^+_1|\lambda\rangle ,$
where $|\lambda\rangle$ belong to the Fock space generated by
$a_\mu^+ , a_D^+$ and depends on the space - time coordinates.

After the substitution of equation (\ref{vector}) into the lagrangian
(\ref{L12})
and integration with respect to $\eta_0$ the lagrangian is expressed only
via the fields $|S_i \rangle$
\begin{eqnarray}\label{MM1}
&&L=-\langle S_1|{\tilde L^0} |S_1\rangle +
 \langle S_2|\tilde L^0|S_2\rangle + \nonumber \\
&&\langle S_1| \tilde L^{1+} |S_3 \rangle + \langle S_3| \tilde L^1
|S_1\rangle
- \langle S_3||S_3\rangle
-  \\
&&\langle S_2| \tilde L^1 |S_3\rangle - \langle S_3| \tilde
L^{1+} |S_2\rangle \nonumber,
\end{eqnarray}
with the following notations used
\begin{equation}
\tilde L^0 = - {p_{\mu}}^2 +m^2
\end{equation}
\begin{equation}
\tilde L^1 = L^1 - m a_D
\end {equation}
\begin{equation} \nonumber
\tilde L^+_1 = L^{1+} -a^+_D m,
\end {equation}
The corresponding equations of motion are:
\begin{eqnarray}
&&\tilde L_0 |S_1 \rangle -\tilde L^+_1 |S_3\rangle=0, \nonumber \\
\label{s2}
&&{\tilde L}_0|S_2\rangle-\tilde L_1 |S_3\rangle=0,
\nonumber \\
&&|S_3\rangle-\tilde L_1 |S_1\rangle + \tilde L^+_1
|S_2\rangle=0.
\end{eqnarray}
While the gauge transformation law (\ref{TR}) in the component form looks
as
follows
\begin{equation}
\delta |S_1\rangle = \tilde{L}_1^{+}{\mid}\lambda\rangle,
\end{equation}
\begin{equation}
\delta|S_2 \rangle = {\tilde{L}_1} {\mid}\lambda\rangle,
\end{equation}
\begin{equation}
\delta |S_3 \rangle=\tilde{L}_0 {\mid}\lambda\rangle.
\end{equation}
Then using the gauge transformations
one can show \cite{PT1}, that the fields $|S_2 \rangle$ and $S_3 \rangle$
as well as $a_D^+$ dependence in $|S_1 \rangle$ can be gauged away.
Finally one obtains conditions
\begin{equation}
(-p^2_\mu  +m^2)|S_1\rangle =
L^1|S_1\rangle = 0
\end{equation}
 as the result of the equations of motion.

Alternatively, one can first gauge away field $|S_2\rangle$ and
$a_D^+$
dependence in fields $|S_1\rangle$ and $|S_3 \rangle$, then express
the field $|S_3 \rangle$ using its own equation of motion
in terms $|S_1 \rangle$ and put back into the lagrangian (\ref{MM1})
to obtain
\begin{equation} \label{MM2}
L =  -\langle S_1|-p_{\mu}^2 + (p_\mu a^+_\mu)(p_\nu a_\nu)
+m^2 |S_1 \rangle \nonumber
\end{equation}
The  lagrangian (\ref{MM2}) (or (\ref{MM1}))  describes all spins from
zero to infinity with mass equal to $m$.
 Due to the
luck of the tracelessness constraint, each spin $n$ in the state $|S_1
\rangle$
 is followed also by spins
$n-2,\; n-4\; ...$ on this level.

So, in this Section we demonstrated the method of BRST construction in the
presence of two second class constraints $[L^1,L^{1+}]= -p_\mu^2=L^0-m^2$.
The price for this was  the introduction of additional creation and
annihilation operators $a^+_D$ and $a_D$ in the Fock space with
corresponding modification of the representation for the basic operators
 $L^1,L^{1+}$. In the next Section the method will be generalized
for more complicated systems of constraints.

\section{IRREDUCIBLE MASSLESS CASE}
\subsection{BRST construction}
The fields we are going to describe using the approach given
in the previous Section correspond to the following
Young tableaux
%\vspace{1cm}\\

\begin{eqnarray}
\label{young}
\begin{picture}(65,30)
\unitlength=1.5mm
\put(0.5,10){\line(1,0){45.2}}
\put(0.5,5){\line(1,0){45.2}}
\put(0.5,0){\line(1,0){35}}
%\put(-20,5){\line(1,0){45}}
%\put(-20,0){\line(1,0){30}}
\put(.5,0){\line(0,1){10}}
\put(5.5,0){\line(0,1){10}}
\put(10.5,0){\line(0,1){10}}
%\put(5,0){\line(0,1){10}}
%\put(10.2,0){\line(0,1){10}}
\put(15,0){\line(0,1){10}}
\put(20,0){\line(0,1){10}}
\put(25,0){\line(0,1){10}}
\put(30,0){\line(0,1){10}}
\put(35.2,0){\line(0,1){10}}
\put(40,5){\line(0,1){5}}
\put(45.2,5){\line(0,1){5}}
\put(1,7){$\mu_1$}
\put(6,7){$\mu_2$}
\put(41,7){$\mu_{n_1}$}
\put(1,2){$\nu_1$}
\put(6,2){$\nu_2$}
\put(31,2){$\nu_{n_2}$}
\multiput(12.5,7.5)(5,0){6}{\circle*{.5}}
\multiput(12.5,2.5)(5,0){4}{\circle*{.5}}
\end{picture}
\end{eqnarray}
and have the mass equal to zero i.e., $L^0 = -p_\mu^2$.
 Constraints describing these fields are
those given in  the first section, for $i,j=1,2$, namely $E^\alpha
=(L^{ij}, T)$,
$E^{-\alpha}=(L^{+ij},T^+)$ and ${\cal{L}}^A = (L^0, L^{i},L^{+i})$.

The operators $E^{\pm \alpha}$ and $H^i$ form the algebra of group
$SO(3,2)$,
which can be written in the compact form as
\begin{eqnarray}
\label{commutator}
&&\left[{H}^i,{E}^\alpha\right]=\alpha(i) {E}^\alpha, \nonumber\\
&&\left[{E}^\alpha,{E}^{-\alpha}\right]=\alpha^i {H}^i, \nonumber \\
&&\left[{E}^\alpha,{E}^{\beta}\right]=N^{\alpha\beta}
{E}^{\alpha+\beta},
\end{eqnarray}
The corresponding nilpotent BRST charge  for this subsystem of constraints
can be  constructed as follows \cite{PT3}.

First we construct the auxiliary representations of generators of
the $SO(3,2)$ group
using the Verma module
after introduction of the additional creation and annihilation
operators $\left[ b_I,b^{+}_J \right] = \delta_{IJ}$,
$I,J=1,...,4$
 The number of
the oscillators is equal to the number of positive roots of the
algebra $SO(3,2)$ and the vector $|\Phi\rangle$
depends also on the creation operators $b^+_I$.

Namely
let us introduce the vector in the  space of Verma module
\begin{eqnarray}\label{b1}
&&{\left|n_1,n_2,n_3,n_4 \right \rangle}_V = \\
&&(L^{11 +})^{n_1}(L^{12+})^{n_2}(L^{22+})^{n_3}
 (T^+)^{n_4}{|0 \rangle}_V \nonumber
\end{eqnarray}
where
 $n_i \in N$ and $E^{\alpha}{|0 \rangle}_V = 0$.
The corresponding vector in the Fock space generated by the creation and
 operators  $b_I^+$ is
  \begin{eqnarray}\label{b2}
&&\left|n_1,n_2,n_3,n_4 \right\rangle = \\
&&(b^+_1)^{n_1}(b^+_2)^{n_2}(b^+_3)^{n_3} (b^+_4)^{n_4}|0 \rangle.
\nonumber
\end{eqnarray}
Mapping the vectors in the Verma module onto
the vectors in the Fock space one obtains the representations of the
generators of $SO(3,2)$ (see \cite{B} for the general construction)
algebra in terms of $b_I$, $b_I^+$
and constant parameters $h^i$, which characterize the highest weight
representation $H_i|0 \rangle_V=h_i|0 \rangle_V$.
However since the vectors in the Fock space form orthogonal basis and the
corresponding vectors in Verma module do not, the correspondence between
this
two spaces is incomplete. In order to establish the complete
correspondence
one has to modify the scalar product in the Fock space under the condition
\begin{equation} \label{def}
\langle\Phi_1 |K|\Phi_2\rangle={}_V\langle\Phi_1 |\Phi_2\rangle_V.
\end{equation}
with he Kernel operator $K$
\begin{equation}
K = Z^+ Z,
\end{equation}
\begin{eqnarray}
Z\!\!\!\!&=&\!\!\!\! \sum_{n_i}
(L^{11 +})^{n_1}(L^{12+})^{n_2}(L^{22+})^{n_3}
 (T^+)^{n_4}{|0 \rangle}_V \!\times \nonumber \\
\!\!\!\!&&\!\!\!\!\langle 0|(b_1)^{n_1}(b_2)^{n_2}(b_3)^{n_3} (b_4)^{n_4}
\frac{1}{\prod n_i!}.
\end{eqnarray}
Next define
\begin{equation} \label{sum}
{\cal H}^i={H}^i + \tilde {H^i}_{aux.} +  h^i ,\quad
\end{equation}
\begin{equation}
{\cal E}^{\pm \alpha}= {E}^{\pm \alpha} + E_{aux.}^{ \alpha}(h),
\end{equation}
where we have explicitly extracted the dependence on parameters $h_i$
in the auxiliary representations of Cartan generators.
The corresponding nilpotent BRST charge
with no $H^i$ dependence has the form
\begin{eqnarray} \label {TTT}
&&\tilde{Q_1}=\sum_{\alpha>0}\left(\eta_\alpha {\cal E}^{-\alpha}+
\eta_{-\alpha}{\cal E}^{\alpha}\right)- \nonumber \\
&&\frac{1}{2}\sum_{\alpha\beta}N^{\alpha\beta}
\eta_{-\alpha}\eta_{-\beta}\cal{P}_{\alpha+\beta}
\end{eqnarray}
where, as a consequence of the procedure of dimensional reduction
described
in the previous Section the parameters $h^i$ have to be substituted by the
expressions
\begin{equation}
-\pi^{i}\!=\!
-{ H}^{i}\! -\tilde{H}^{i}_{aux}\!-\!
\sum_{\beta>0}\beta(i)\left(\eta_{\beta}\cal{P}_{-\beta}\!-\!
\eta_{-\beta}\cal{P}_\beta \right)\!.
\end{equation}

Indeed, after the construction of the standard nilpotent BRST charge $Q_1$
treating ${\cal E}^{\pm \alpha}$ and ${\cal H}^i$
 as  first class constraints, one
considers the auxiliary phase space $(x_i, p^i)$ with $p^i = h^i$.
After that one makes the similarity transformation
$\tilde{Q_1} = e^{i \pi^i x_i} Q_1 e^{-i \pi^i x_i}$
analogous to the one considered in the previous section.
This transformation removes the dependence
on the $\eta_i$ ghost variables, which correspond to the Cartan generators
${\cal H}^i$.  Then the ${\cal P}_i$ independent part of the transformed
BRST charge is nilpotent and equals to (\ref{TTT}).

The inclusion of the constraints $\cal{L}^A$ $\equiv (L^0, L^i, L^{i+})$
into the total BRST charge $Q$ is trivial, namely
\begin{equation} \label{BRST}
\tilde{Q}= \tilde{Q_1} + \tilde{Q_2}
\end{equation}
where
\begin{eqnarray}
&&\tilde{Q_2} = \eta_0 L^0 + \eta_i L^{i +} + \eta^{+}_iL^i -
\eta^{ +}_i \eta_i \cal{P}_0
+ \nonumber \\
&&\sum_{\alpha > 0, A,B}(\eta_A \eta^{+}_\alpha \cal{P}_B C_{- \alpha,
A}^B
+  \eta_A \eta_{\alpha} \cal{P}_B C_{ \alpha, A}^B)
\end{eqnarray}
in self explanatory notations. This completes the procedure
of constructing nilpotent BRST charge for our system.

The BRST invariant lagrangian  can be written as
\begin{equation} \label{L}
- L = \int d \eta_0 \langle \chi| K \tilde{Q} | \chi \rangle,
\end{equation}
being invariant under the gauge transformations
\begin{equation} \label{G}
\delta | \chi \rangle = \tilde{Q} | \Lambda \rangle
\end{equation}
Following the lines of the previous section, after the integration over
the ghost variables and elimination of auxiliary fields using the
equations of motion and the BRST gauge transformations one arrives
to the final expression for the lagrangian \cite {PT4} describing all
massless
irreducible representations of the Poincare group with the corresponding
Young tableaux having two rows
\begin{eqnarray} \label{F}    \nonumber
\!\!\!\!\!\!&&-L = \nonumber \\
\!\!\!\!\!\!&&\langle S_1| L^{0}  -  L^{+1}  L^{1}  -
       L^{+2} L^{2}  -
       L^{+1}  L^{+1}  L^{11}- \nonumber \\
\!\!\!\!\!\!&& L^{+1}  L^{+2}  L^{12}   -
        L^{+2}  L^{+2}  L^{22}   -
      2   L^{+11}  L^{0}  L^{11}
- \nonumber \\
\!\!\!\!\!\!&&L^{+11}  L^{1}  L^{1}
 - L^{+12}  L^{0}  L^{12}   -
        L^{+12}  L^{1}  L^{2}- \nonumber \\
 \!\!\!\!\!\!  && 2   L^{+22}  L^{0}  L^{22}   -
        L^{+22}  L^{2}  L^{2} - \nonumber \\
\!\!\!\!\!\!&&  L^{+1}  L^{+11}  L^{1}  L^{11}
- L^{+1}  L^{+12}  L^{2}  L^{11}+ \nonumber \\
\!\!\!\!\!\!  && L^{+1}  L^{+22}  L^{1}  L^{22}
 - L^{+1}  L^{+22}  L^{2}  L^{12}   -\nonumber \\
\!\!\!\!\!\!     && L^{+2}  L^{+11}  L^{1}  L^{12}
+ L^{+2}  L^{+11}  L^{2}  L^{11}-  \nonumber \\
\!\!\!\!\!\!&&L^{+2}  L^{+12}  L^{1}  L^{22}   -
        L^{+2}  L^{+22}  L^{2}  L^{22}   + \nonumber \\
\!\!\!\!\!\!        &&L^{+1}  L^{+1}  L^{+22}  L^{11}  L^{22}
-  L^{+1}  L^{+2}  L^{+12}  L^{11}  L^{22}   +
      \nonumber \\
\!\!\!\!\!\!&&  L^{+2}  L^{+2}  L^{+11}  L^{11}  L^{22}   +
      3   L^{+11}  L^{+22}  L^{0}  L^{11}  L^{22} +\nonumber \\
\!\!\!\!\!\!&& L^{+11}  L^{+22}  L^{1}  L^{1}  L^{22}
- L^{+11}  L^{+22}  L^{1}  L^{2}  L^{12}    + \nonumber \\
\!\!\!\!\!\!&&  L^{+11}  L^{+22}  L^{2}  L^{2}  L^{11} +  \nonumber \\
\!\!\!\!\!\!&& L^{+1}  L^{+11}  L^{+22}  L^{1}  L^{11}  L^{22}   +
\nonumber \\
\!\!\!\!\!\!&& L^{+2}  L^{+11}  L^{+22}  L^{2}  L^{11}  L^{22} |S_1\rangle
\end{eqnarray}
where  the field $|S_1 \rangle$ is constrained as
\begin{equation} \label{USL1}
T|S_1\rangle=0
\end{equation}
\begin{eqnarray} \label{USL2}
L^{\{ij}L^{kl \}} |S_1\rangle=0
\end{eqnarray}
In particular, the equation (\ref{USL1}) leads also to the following
property of the
basic field
\begin{eqnarray}\label{confsym}
&&\Phi_{\mu_1\mu_2\cdots\mu_{n}, \nu_1\nu_2\cdots\nu_{n}}(x)= \nonumber \\
&&(-1)^n \Phi_{\nu_1\nu_2\cdots\nu_{n}, \mu_1\mu_2\cdots\mu_{n}}(x),
\end{eqnarray}
when the numbers of indices $\mu$ and $\nu$ coincide.

The lagrangian (\ref{F}) is invariant under the transformations
\begin{equation} \label{cal}
\delta |S_1 \rangle = (L^{1+} +L^{2+}T)|\lambda\rangle
\end{equation}
with the parameter of gauge transformations $|\lambda \rangle$
constrained as follows
\begin{equation}\label{constraintsforlambda}
L^{ij}|\lambda\rangle = T^2 |\lambda\rangle =0
\end{equation}
obviously the constraints on the  basic field (\ref{USL1}) -(\ref{USL2})
are also invariant under the gauge transformations (\ref{cal}).

Let us note, that neglecting the $a^{2+}_\mu$ dependence
in the field $|S_1 \rangle$ one obtains
the lagrangian given in \cite{F} for irreducible massless higher spin
fields.

\subsection{Examples}
Let us construct
the explicit form of  the lagrangians for some
simple Young tableaux which correspond to lower orders in the expansion
of the field $|S_0 \rangle$.

$\bullet$
\begin{picture}(20,20)
\unitlength=0.5mm
\put(0,7.5){\line(1,0){5}}
\put(0,2.5){\line(1,0){5}}
\put(0,-2.5){\line(1,0){5}}
\put(0,-2.5){\line(0,1){10}}
\put(5,-2.5){\line(0,1){10}}
\end{picture}:
$|S_1\rangle=\Phi_{\mu,\nu}(x)a^{1+}_{\mu}a^{2+}_{\nu}|0\rangle$\vspace{5mm}\\
The lagrangian  (\ref {F}) for this antisymmetric field
in terms of the field strength
$F_{\mu\nu\rho}=\partial_\mu\Phi_{\nu,\rho}+
\partial_\nu\Phi_{\rho,\mu}+\partial_\rho\Phi_{\mu,\nu}$
has the standard form
\begin{equation}\label{LS11}
  L=\frac{1}{3}F_{\mu\nu\rho}^2
\end{equation}
and is invariant under the  gauge transformations
\begin{equation}\label{G11}
  \delta\Phi_{\mu,\nu}(x)=\partial_\mu\lambda_\nu(x)-
\partial_\nu\lambda_\mu(x).
\end{equation}

$\bullet$
%\item
\begin{picture}(20,20)
\unitlength=0.5mm
\put(0,7.5){\line(1,0){10}}
\put(0,2.5){\line(1,0){10}}
\put(0,-2.5){\line(1,0){5}}
\put(0,-2.5){\line(0,1){10}}
\put(5,-2.5){\line(0,1){10}}
\put(10,2.5){\line(0,1){5}}
\end{picture}:
 $|S_1\rangle=\Phi_{\mu\nu,\rho}(x)a^{1+}_{\mu}a^{1+}_{\nu}
a^{2+}_{\rho}|0\rangle$\vspace{5mm}\\
The symmetry   with respect to the first two indices
$\Phi_{\mu\nu,\rho}=\Phi_{\nu\mu,\rho}$
which is guaranteed by the construction and the condition
(\ref{USL1}) lead to the following property of
the field  $\Phi_{\mu\nu,\rho}$:
\begin{equation}\label{S21}
 \Phi_{\mu\nu,\rho}+\Phi_{\rho\nu,\mu}+\Phi_{\mu\rho,\nu}=0.
\end{equation}
 The
lagrangian (\ref{F}) for the third rank tensor
field $\Phi_{\mu\nu,\rho}$  can
be written in the form
\begin{eqnarray}\label{L21}
&&  L=2 \Phi_{\mu\nu,\rho}\partial^2_\sigma \Phi_{\mu\nu,\rho}-
3\Phi_{\mu\mu,\rho}\partial^2_\sigma \Phi_{\nu\nu,\rho}- \nonumber \\
&&4\Phi_{\mu\nu,\rho}\partial_\mu\partial_\sigma \Phi_{\sigma\nu,\rho}-
2 \Phi_{\mu\nu,\rho}\partial_\rho\partial_\sigma
\Phi_{\mu\nu,\sigma} +\nonumber\\
&&6\Phi_{\mu\nu,\rho}\partial_\mu\partial_\nu \Phi_{\sigma\sigma,\rho}+
3\Phi_{\mu\mu,\rho}\partial_\rho\partial_\sigma \Phi_{\nu\nu,\sigma},
\end{eqnarray}
and is invariant under the gauge transformations
\begin{eqnarray}\label{G21}
&&  \delta\Phi_{\mu\nu,\rho}(x)=\partial_\mu\lambda_{\nu,\rho}(x)+
\partial_\nu\lambda_{\mu,\rho}(x)- \nonumber \\
&&
\partial_\rho\lambda_{\mu,\nu}(x)-
\partial_\rho\lambda_{\nu,\mu}(x),
\end{eqnarray}
the gauge transformation parameter being traceless
$\lambda_{\mu,\mu}=0$.

$\bullet$ %\item
\begin{picture}(20,20)
\unitlength=0.5mm
\put(0,7.5){\line(1,0){10}}
\put(0,2.5){\line(1,0){10}}
\put(0,-2.5){\line(1,0){10}}
\put(0,-2.5){\line(0,1){10}}
\put(5,-2.5){\line(0,1){10}}
\put(10,-2.5){\line(0,1){10}}
\end{picture}:
 $|S_1\rangle=\Phi_{\mu\nu,\rho\sigma}(x)a^{1+}_{\mu}a^{1+}_{\nu}
a^{2+}_\rho a^{2+}_\sigma |0\rangle$\vspace{5mm}\\
The field $\Phi_{\mu \nu, \rho \sigma}$
can be related to the field $C_{\mu \nu, \rho \sigma}$
which has  symmetries  of the Weyl tensor
\begin{equation}
C_{\mu \nu, \rho \sigma} = - C_{\nu \mu, \rho \sigma}=-C_{\mu \nu, \sigma
\rho},
\end{equation}
\begin{equation}
 C_{\mu \nu, \rho \sigma} = C_{\rho \sigma, \mu \nu},
\end{equation}
in terms of relations
\begin{eqnarray}
\Phi_{\mu \nu, \rho \sigma} &=& \frac{1}{4}
(C_{\mu \rho, \nu \sigma} + C_{\mu \sigma, \nu \rho})\\
C_{\mu \rho, \nu \sigma} &=& \frac{4}{3}(\Phi_{\mu \nu, \rho \sigma}-
\Phi_{\rho \nu, \mu \sigma}).
\end{eqnarray}

Strictly speaking $C_{\mu \nu, \rho \sigma}$  is not a Weyl tensor because
its traces
do not vanish. However,
one can obtain  from (\ref{F})  the following lagrangian
\begin{eqnarray}\label{W}
\!\!\!\!\!\!\!\!&&  L=-\frac{1}{2}
C_{\mu\rho,\nu \tau}\partial^2_\sigma C_{\mu\rho,\nu\tau}
-\frac{1}{2}
C_{\mu\tau,\nu \rho}\partial^2_\sigma C_{\mu\rho,\nu\tau} + \nonumber \\
\!\!\!\!\!\!\!\!&& 3
C_{\mu\rho,\nu \rho}\partial^2_\sigma C_{\mu\tau,\nu\tau}
 - \frac{3}{4}
C_{\mu\nu,\mu \nu}\partial^2_\sigma C_{\rho\tau,\rho\tau} + \nonumber \\
\!\!\!\!\!\!\!\!&&2
C_{\mu\rho,\nu \sigma}\partial_\mu \partial_\tau
 C_{\tau\rho,\nu\sigma}
+2
C_{\mu\rho,\nu \sigma}\partial_\mu \partial_\tau
 C_{\tau\sigma,\nu\rho}- \nonumber \\
\!\!\!\!\!\!\!\!&&
6
C_{\mu\rho,\nu \sigma}\partial_\mu \partial_\nu
 C_{\tau\rho,\tau\sigma}
-6
C_{\mu\rho,\nu \rho}\partial_\mu \partial_\sigma
 C_{\sigma\tau,\nu\tau} + \nonumber \\
\!\!\!\!\!\!\!\!&&3
C_{\mu\rho,\nu \rho}\partial_\mu \partial_\nu
 C_{\sigma\tau,\sigma\tau}
\end{eqnarray}
invariant under the gauge transformations
\begin{eqnarray}
\!\!\!\!\!\!\!\!&&\delta C_{\mu\nu,a b}(x)=\partial_\mu
\lambda_{\nu,ab}(x)
- 2 \partial_a \lambda_{\mu,\nu b}(x) - \nonumber \\
\!\!\!\!\!\!\!\!&& (\mu \leftrightarrow a, \nu \leftrightarrow b)
\end{eqnarray}
\begin{equation}
\lambda_{\mu,\mu \nu}= \lambda_{\mu,\nu \nu}=0, \quad
\lambda_{\mu,\nu \rho}=\lambda_{\mu, \rho \nu}
\end{equation}
\begin{equation}
\lambda_{\mu,\nu\rho}+ \lambda_{\nu,\rho\mu}+\lambda_{\rho,\mu\nu}=0
\end{equation}
where the symmetrization over couples of Greek and Latin
indices is assumed.
It can be shown that  the vanishing of all
traces
of this tensor on mass shell follows from this lagrangian . Therefore one
can conclude
 that the lagrangian (\ref{W}) consistently
describes the free field theory of the Weyl tensor.

\section{DISCUSSION}

The approach we have described can be straightforwardly generalized to the
particles which belong to an arbitrary representations of the Poincare
group as well.
Another challenging problem is construction of the theory of interacting
higher spin fields.
It is known that the particles
with the higher spins can propagate through the background
having the constant curvature, in particular through the AdS space
(see \cite{MV} and the references therein),
 as well as interact with the constant electromagnetic
field or symmetrical Einstein spaces  \cite{KL}.
The utilization of the technique of the Supersymmetric Quantum
Mechanics leads also to the description of the particle with spin 2
on the background of the constant curvature \cite{KY},
and on the background being real ``Kahler -- like" manifold
\cite{DP}.
In the highlight of these developments
it seems  interesting to generalize this technique also
for the description of the interaction of
higher spin fields
with some gravitational background.
The inclusion of gravitational background will obviously lead
to the modification of the system of
constraints presented in the BRST
charge. The problem of constructing of the nilpotent BRST charge for
these kinds of physical systems can in
turn reveal an allowed
types of gravitational backgrounds where  the higher spin
fields can propagate consistently.

\vspace{1.3cm}

\noindent {\bf Acknowledgments}
~We ~are ~grateful to G. Thompson ~for ~bringing the ~reference \cite{TH}
to ~our ~attention.
~This ~investigation ~has ~been ~supported
~by ~the
~grant ~of ~the ~Committee ~for collaboration between Czech Republic and
JINR.
The work of \v{C}.B. was supported by the grant of the Ministry
of Education of Czech Republic VZ/400/00/18. The work of A.P. and M.T.
was supported
 in part by the
Russian Foundation of Basic Research,
grant 99-02-18417 and the joint grant RFBR-DFG
99-02-04022.

\end{document}